# An AI-enabled dual-hormone model predictive control algorithm that delivers insulin and pramlintide

**Peter G. Jacobs*┼, Wade Hilts*, Robert Dodier*, Joseph Leitschuh*, Jae H. Eom**, Deborah Branigan**, Forrest Ling**, Matthew Howard**, Clara Mosquera-Lopez*, Leah Wilson****

**Artificial Intelligence for Medical Systems Lab, Department of Biomedical Engineering, Oregon Health & Science University, Portland OR 97239 (jacobsp@ohsu.edu)*

*┼Artificial Intelligence for Medical Systems Lab, College of Bioengineering, Oregon State University, Portland, OR 97239*

*** Harold Schnitzer Diabetes Health Center, Oregon Health & Science University, Portland OR 97239*

**Abstract**: Current closed-loop insulin delivery algorithms need to be informed of carbohydrate intake disturbances to perform well. This can be a burden on people using these systems. Pramlintide is a hormone that delays gastric emptying, which can enable insulin kinetics to align more closely with the kinetics of carbohydrate absorption. Integrating pramlintide into an automated insulin delivery system (AID) can be helpful in reducing the postprandial glucose excursion and may be helpful in enabling fully-closed loop whereby meals do not need to be announced. We present an AI-enabled dual-hormone model predictive control (MPC) algorithm that delivers insulin and pramlintide without requiring meal announcements that uses a neural network to automatically detect and deliver meal insulin. The MPC algorithm includes a new pramlintide pharmacokinetics and pharmacodynamics model (PK/PD) that was identified using data collected from people with type 1 diabetes undergoing a meal challenge. Using a simulator, we evaluated the performance of various pramlintide delivery methods and controller models, as well as the baseline insulin-only scenario. Meals were automatically dosed using a neural network meal detection and dosing (MDD) algorithm. The primary outcome measure was the percent time glucose is in time in target range (TIR: 70-180 mg/dL). Results show that delivering pramlintide at a fixed ratio of 6 mcg pramlintide : 1 u insulin using an MPC that is aware of the pramlintide achieved the most significant improved TIR compared with an insulin-only MPC (91.6% vs. 64.1%). Delivering pramlintide as either a fixed ratio or independent control was better than delivering basal pramlintide at a constant rate, indicating the benefit of the MDD algorithm. There was not an advantage of independent control of insulin and pramlintide compared with insulin and pramlintide delivered as a fixed ratio. Preliminary real-world results from a human subject further indicate the benefit of pramlintide delivered as a fixed ratio with insulin.

## 1. INTRODUCTION

### *1.1 Handling meals in automated insulin delivery*

Many commercial automated insulin delivery (AID) systems are *hybrid AIDs,* meaning that users need to estimate their carbohydrates prior to meals and announce these to the system to avoid excessive high glucose after meals. People with T1D using AID systems oftentimes forget to dose their meal insulin or misestimate their carbohydrates. Newer commercial AIDs such as the Medtronic 780g can automatically dose correction insulin when meals are detected (Shalit 2023). In a prior outpatient study using an AI-enabled meal detection and dosing algorithm AID, we showed that when a robust AI-enabled meal detection and dosing algorithm (RAP) was used to dose for meals in a full closed-loop (FCL) environment, the time above range (>180 mg/dL) could be reduced by 10.8% compared with a standard model predictive control (MPC) algorithm that did not include automated meal dosing (Mosquera-Lopez 2023). The algorithm detected and dosed for 83.3% of meals on average 25.9 minutes after the meal.

While automated meal detection and delivery can be effective at modestly improving glucose outcomes when people forget to announce meals using a hybrid AID, because of the delayed kinetics of subcutaneously delivered insulin relative to the kinetics of carbohydrate absorption (Miles, et al. 1995), current insulin alone solutions are not effective for enabling full automation in response to meals. Integrating newer ultra-fast acting insulin such as Lyumjev in combination with a highly accurate pump that delivers precise insulin doses with automated meal detection and dosing could improve FCL (Leohr, et al. 2024).

### *1.2 Integrating pramlintide into AID*

Integrating pramlintide into a dual hormone hybrid closed loop system may further improve time in range (TIR, 70-180 mg/dL) during the daytime as shown by Haidar and colleagues (2020), who found that TIR could be increased from 74% (SD 18%) to 84% (SD 13%) when incorporating pramlintide dosed with insulin prior to meals. Pramlintide is a synthetic form of amylin which is released with insulin from the beta cell in normal physiology. People with T1D lose both insulin and amylin secretion due to autoimmune beta cell destruction (Kruger, et al. 1999). Pramlintide substantially blunts rate of appearance of glucose after a meal as shown by Riddle et al. (2018) by suppressing glucagon production to prevent hepatic glucose release (Nyholm, et al. 1999) and delaying gastric emptying (Vella, et al. 2002) via central nervous system effects. The slowing of gastric emptying by pramlintide controls the influx of glucose from the meal to better match the relatively slower kinetics of subcutaneous insulin for disposal of glucose. Kong et al. (1997) demonstrated that after a meal in adults with T1D, gastric emptying was ~10% at 40 minutes after the meal, enabling time for prandial delivery of insulin to reduce postprandial excursions.

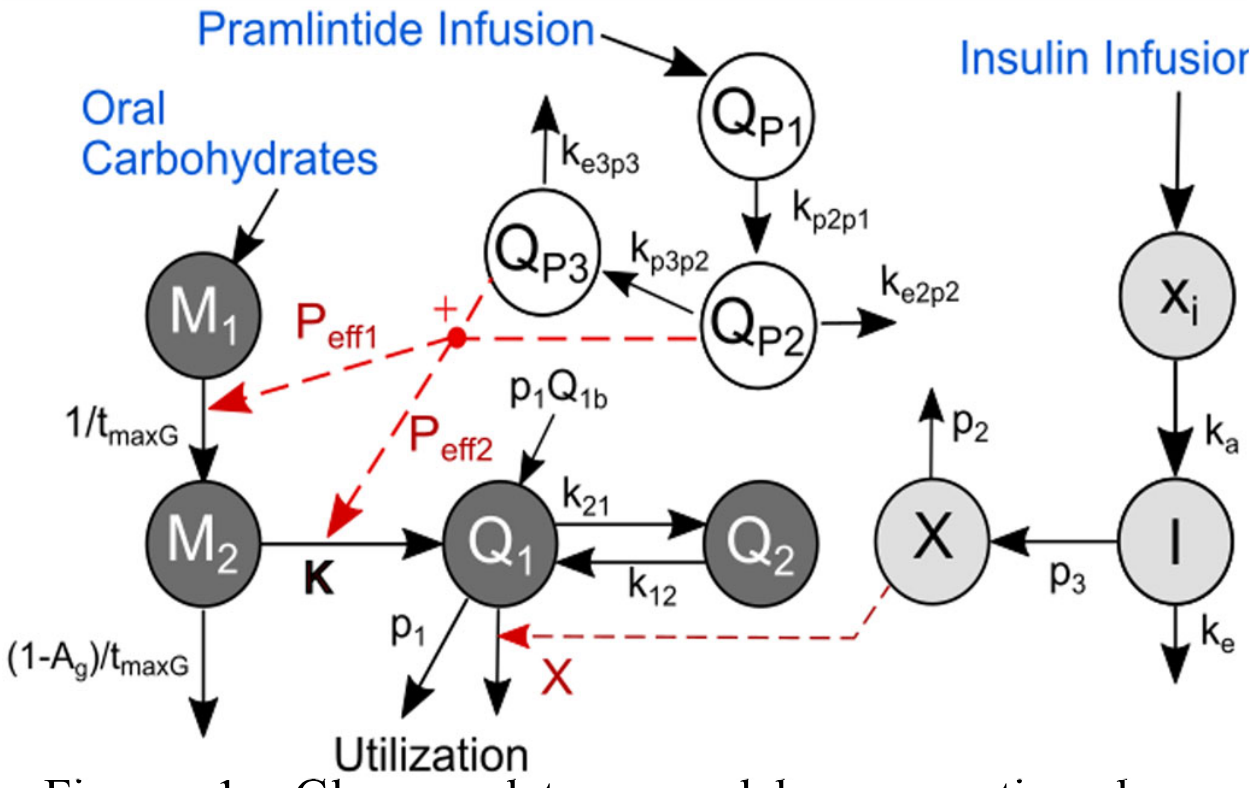

Figure 1: Glucoregulatory model representing how pramlintide impacts delayed gastric emptying.

## 2. METHODS

### *2.1 Introducing a new pramlintide PK/PD model*

We used data collected from Adocia (Andersen et al. 2010) to identify a new PK/PD model of pramlintide.

The dataset that we used to identify the model was collected during a meal tolerance test whereby participants came into a clinic and consumed a meal while receiving a proprietary co-formulation of insulin and pramlintide. Each participant received an 88 g carb meal with 45 mcg pramlintide and 7.5 U insulin. Blood metabolite data including glucose, insulin, and pramlintide (both intact pramlintide and an active metabolite, des-lys pramlintide) was collected for the next several hours. Models were fit using Hamiltonian Monte Carlo methods using Stan (2020). The model architecture that we selected is shown in Figure 1. States for the model are in Table 1 and parameters for that model are in Table 2.

### *2.2 Model used within the MPC*

We previously described the design of a single-hormone MPC and a dual-hormone MPC (insulin and glucagon) in a prior IFAC publication (Resalat 2017). In that model we used a minimal insulin dynamics model and a 2-compartment insulin kinetics model to represent insulin PK/PD (Cobelli 1999, Kobayashi 1983). In the current manuscript, we extend the single-hormone insulin-only MPC described in that manuscript to include a new model that models how pramlintide impacts a delay on gastric emptying. The new model used in the MPC is shown in Figure 1. In this model, oral carbohydrates are consumed and move from the gut, $M_1$, into an unobservable compartment, $M_2$, and finally into plasma, $Q_1$. Subcutaneous insulin is infused into compartment $X_i$ and moves into plasma, I. Subcutaneous pramlintide is infused into compartment $Q_{P1}$ and then moves into blood plasma as intact pramlintide, denoted, $Q_{P2}$, and then as the active metabolite, denoted $Q_{P3}$. Information from our industry partner indicates that the active metabolite has approximately the same physiological effect as intact pramlintide. The effect of pramlintide on the movement of carbohydrates into plasma is represented as the action of P on the gastric emptying coefficient $1/t_{maxG}$.

| State Variable | Description | Units |
|---|---|---|
| $Q_1$ | Plasma Glucose | mg/kg |
| $Q_2$ | Glucose in inaccessible compartment | mg/kg |
| $x_i$ | Subcutaneous insulin | mU/kg |
| $I$ | Plasma insulin concentration | mU/kg |
| $X$ | Effect of insulin in interstitial fluid | no units |
| $M_1$ | First meal compartment | g |
| $M_2$ | Second meal compartment | g |
| $Q_{P1}$ | First pramlintide compartment | pg/kg |
| $Q_{P2}$ | Second pramlintide compartment | pg/kg |
| $Q_{P3}$ | Third pramlintide compartment | pg/kg |

Table 1: Model state descriptions and units.

When a person eats a meal, the carbohydrates ($U_m$) enter the gut ($M_1$) and move into $M_2$ with a rate of ${}^{1}/_{t_{maxG}}$.

$$\dot{M}_1 = U_m - \frac{M_1}{t_{\mathrm{maxG}}\left(\frac{Sf_{P_1}\left(Q_{P_2}+Q_{P_3}\right)}{V_{dP}}+1\right)} \tag{1}$$

$$\dot{M}_2 = \frac{M_1}{t_{\mathrm{maxG}}\left(\frac{Sf_{P_1}\left(Q_{P_2}+Q_{P_3}\right)}{V_{dP}}+1\right)} - \frac{M_2}{t_{\mathrm{maxG}}\left(\frac{Sf_{P_2}\left(Q_{P_2}+Q_{P_3}\right)}{V_{dP}}+1\right)} \tag{2}$$

The percent of the meal utilized is handled by the variable $A_g$ whereby an Ag of 1.0 indicates full meal utilization, and 0.5 indicates a 50% meal utilization. $A_g$ is included in the compartment transfer variable K, which incorporates the person's body weight and converts the carbohydrate amount to the correct units whereby the 180 is a unit conversion term from mmol/kg into mg/kg.

$$K = \frac{5.556 * A_g * 180}{t_{maxG} * weight} \tag{3}$$

The effect of the pramlintide on gastric emptying is given by the sensitivity factors $Sf_{P1}$ and $Sf_{P2}$. We assumed that the intact and metabolite pramlintide act equivalently on the gastric emptying. The concentration of active pramlintide (P) is therefore the sum of the metabolite and the intact pramlintide (dividing by the volume of distribution of pramlintide, $V_P$, to convert mass per mass to mass per volume).

$$P = \frac{Q_{P_2} + Q_{P_3}}{V_P} \tag{4}$$

The kinetics of pramlintide moving from subcutaneous into plasma are given by the following equations.

$$\dot{Q_{P_1}} = U_P - k_{P_2,P_1}\, Q_{P_1} \tag{5}$$

$$\dot{Q_{P_2}} = k_{P_2,P_1}\, Q_{P_1} - k_{e_2,P_2}\, Q_{P_2} - k_{P_3,P_2}\, Q_{P_2} \tag{6}$$

$$\dot{Q_{P_3}} = k_{P_3,P_2}\, Q_{P_2} - k_{e_3,P_3}\, Q_{P_3} \tag{7}$$

The effect of pramlintide on gastric emptying is represented by the pramlintide effect variables $P_{Eff1}$ and $P_{Eff2}$, which are functions of $Sf_{P1}$ and $Sf_{P2}$.

$$P_{Eff1} = \frac{1}{1 + SfP_1 * P} \tag{8}$$

$$P_{Eff2} = \frac{1}{1 + SfP_2 * P} \tag{9}$$

The meal carbohydrates move from $M_2$ into plasma ($Q_1$). Glucose disposal out of $Q_1$ happens at a rate of $p_1$ which represents combined disposal due to brain and kidney uptake.

$$\dot{Q}_1 = (-X - p_1 - k_{21})\, Q_1 + k_{12}\, Q_2 + p_1 Q_{1b} + \frac{K\, M_2}{\frac{P_{Eff1}\left(Q_{P_2}+Q_{P_3}\right)}{V_{dP}}+1} \tag{10}$$

$$\dot{Q}_2 = k_{2,1}\, Q_1 - k_{1,2}\, Q_2 \tag{11}$$

Insulin is delivered ($U_I$) to the subcutaneous space ($X_I$) and moves into the insulin in plasma compartment I with a rate constant $k_a$.

$$\dot{x}_I = U_I - k_a\, x_I \tag{12}$$

$$\dot{I} = \frac{k_a x_I}{V_{dI}} - k_e\, I \tag{13}$$

$$\dot{X} = p_3\, I - p_2\, X \tag{14}$$

| Parameter Symbol | Value | Units |
|---|---|---|
| $p_1$ | 0.0122 | Glucose effectiveness, $min^{-1}$ |
| $Q_1 b$ | 260 | mg/kg |
| $p_2$ | 0.035 | min−1 |
| $k21$ | 0.058 | min−1 |
| $k12$ | 0.0885 | min−1 |
| $k_a$ | 0.026 | min−1 |
| $k_e$ | 0.013 | Insulin elimination from Plasma, $min^{-1}$ |
| $tmaxG$ | 40 | Meal model time constant, min |
| $A_G$ | 0.8 | Meal carb glucose efficiency coefficient |
| $VdI$ | 1.274 | Insulin volume of distribution, L/kg |
| $VdG$ | 1.289 | Glucose volume of distribution, dL/kg |
| $kP_2,P_1$ | 0.036103 | min−1 |
| $kP_3,P_2$ | 0.02693735 | min−1 |
| $ke_2,P_2$ | 0.504307 | min−1 |
| $ke_3,P_3$ | 0.02199495 | min−1 |
| $V_P$ | 104.478 | Volume of distribution for pramlintide, mL/kg |
| $SfP1$ | 0.02831485 | Pramlintide effect coefficient 1 |
| $SfP2$ | 0.0189834 | Pramlintide effect coefficient 2 |
| $T_s$ | 5.0 | Sampling time, min |

Table 2: Model parameter values.

Insulin in plasma (X) acts to mediate disposal of glucose out of $Q_1$ according to Equation 10. Insulin also moves into an unobservable compartment $Q_2$ at a rate $k_{21}$.

*2.3 Linearizing the glucoregulatory model*

The model described above is nonlinear in that states are being multiplied together in the state-space equations (e.g. in Equation 10, $Q_1$ is being multiplied by X). It is simpler and less computationally complex to integrate a linear model into an MPC framework compared with a nonlinear model. We evaluated two different ways to linearize the model, a zero-order hold approach (ZOH) and a first-order hold (FOH) approach. There are two places in the model where a nonlinearity occurs, the impact of pramlintide on the gastric emptying (Equations 1-2) and the impact of insulin and pramlintide on the glucose disposal (Equation 10).

*2.3 Linearizing the glucoregulatory model*

We put the Equations 1-14 into state-space format in which states $X_1$ through $X_{10}$ are $Q_1$, $Q_2$, $X_I$, I, X, $M_1$, $M_2$, $Q_{P1}$, $Q_{P2}$, and $Q_{P3}$, respectively. The state-space representation of the system is then given in Equation 15.

$$\dot{X} = AX + Bu_c + Gu_m + D \quad (15)$$

A is the state transition matrix, X is the vector of state values, B represents the impact of the inputs (pramlintide and insulin) on the change in states. G represents the impact of meals ($u_m$) on the change in states. The vector D represents terms which are constant with respect to the state variables; these are mostly derived by the linearization process.

For ZOH linearization, we set the nonlinear components of the state transition matrix to their current value and presume that they remain constant during the prediction horizon.

For FOH, we start with the original nonlinear ODEs,

$$\dot{X} = F(X) \quad (16)$$

where the function F is the right-hand side of the system of differential equations. Note that F is a nonlinear function mapping $\mathbb{R}^n \to \mathbb{R}^n$, with $n$ being the number of state variables of the model. To obtain $A$ and $d$, we will make a first-order approximation, where $X^* = X(t^*)$ is the point around which the equations are linearized; the linearization time $t^*$ is typically the current time when the next control step is to be computed.

$$\begin{aligned} \dot{X} &= F(X^* + (X - X^*))) \\ \dot{X} &\approx F(X^*) + D_X F(X^*)(X - X^*) \\ \dot{X} &= D_X F(X^*)X - D_X F(X^*)X^* + F(X^*) \end{aligned} \quad (17)$$

where $D_X F$ is the Jacobian matrix of $F$ with respect to $X$, with the $i,j$'th element of the $D_X F$ equal to $dF_i/dX_j$. We can write Equation 17 more briefly as

$$\frac{dX}{dt} = A(X^*)X + d(X^*)$$

$$\begin{aligned} A(X^*) &= D_X F(X^*) \\ d(X^*) &= -D_X F(X^*)X^* + F(X^*) \end{aligned} \quad (18)$$

Note that the terms $A(X^*)$ and $d(X^*)$ are only functions of the approximation point $X^*$, so Equation 18 is a linear inhomogeneous ODE for $X$. The linearized version of the system in equations 1-14 are provided in the Supplementary Materials on arxiv (Jacobs et al. 2025).

*2.5 Single and Dual-hormone MPC architecture*

We followed the same approach to dual-hormone MPC described in a prior publications (Resalat 2016, 2017). The state space representation of the MPC process model is given in Equations 19-20.

$$x_m(k+1) = A_m x_m(k) + B_m u(k) + d_m(k) \quad (19)$$

$$y(k) = C_m x_m(k) \quad (20)$$

where $x_m(k)$ is the state vector, u(k) is the input vector (insulin and pramlintide), and d(k) is the constant term originating from the linearization of the nonlinear interactions between the states using the linearization approaches described above. The linearized process model described above is used to determine the predicted glucose levels over a prediction horizon of 300 minutes (NP), which are then compared with a reference trajectory. When the person's glucose is above a target glucose level (e.g. 110 mg/dL), then the reference trajectory is a straight line from the current glucose level to the target glucose level across the prediction horizon. When glucose is less than the target glucose level, the reference trajectory exponentially approaches the target as described in Resalat et al (2017). The complete description of the MPC methods are in Resalat et al. (2016).

*2.6 Meal detection and dosing*

We used a previously described meal detection and dosing (MDD) algorithm described in Mosquera-Lopez et al. (2023) that includes a neural network to estimate the probability of a meal and the meal size. The MDD algorithm dosed the amount of insulin that would optimize the benefit using a digital twin to replay various meal doses in response to the detected meal. The final amount of insulin was limited by a safety layer that would prevent delivery of insulin that would bring the person's

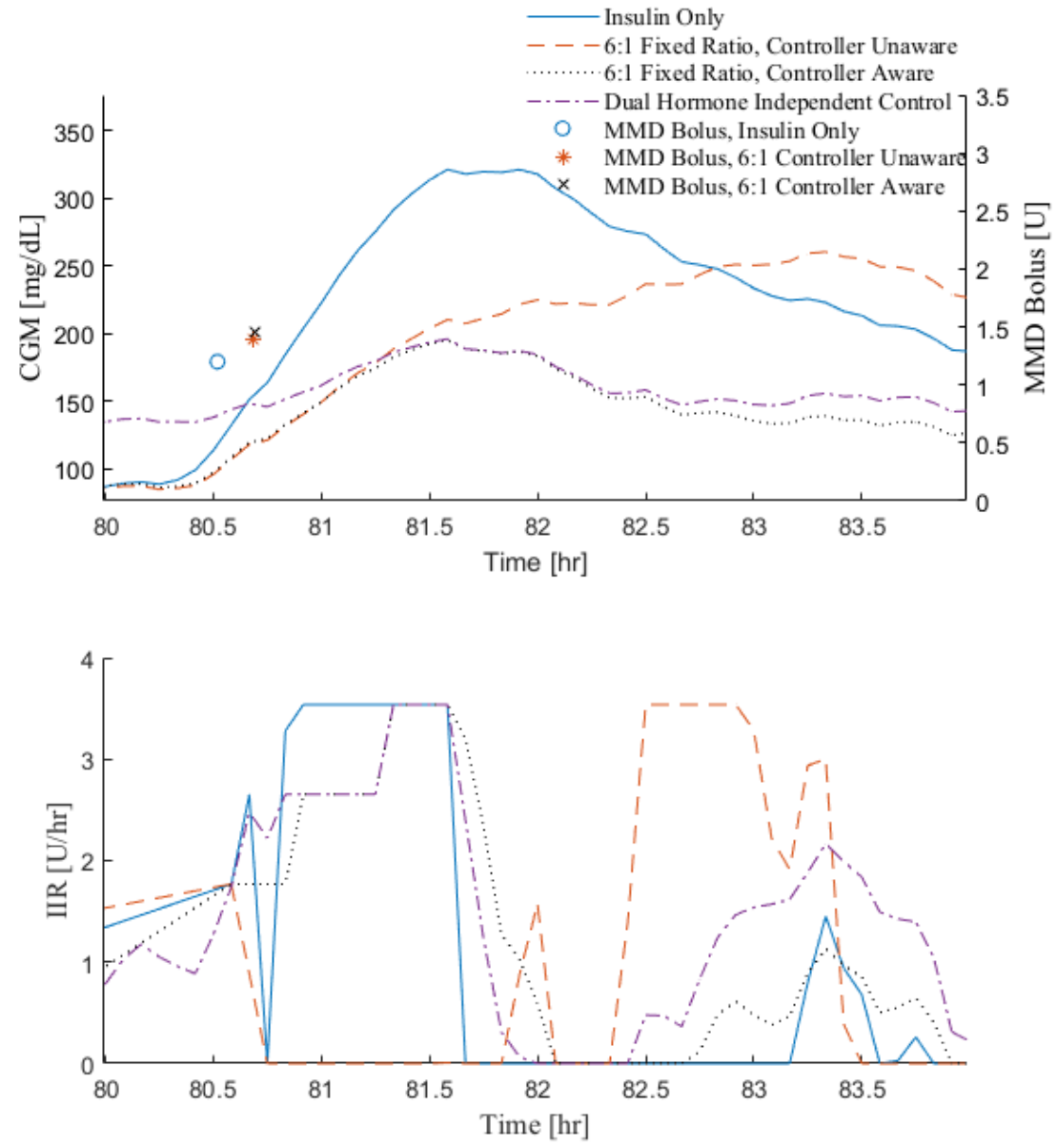

Figure 2: Results from *in silico* subject consuming the same meal under four different scenarios, (1) insulin only AID, (2) pramlintide dosed as 6:1 fixed ratio with insulin with an AID unaware of the pramlintide, (3) pramlintide dosed as a 6:1 fixed ratio with MPC aware of the pramlintide in the process model, and (4) dual-hormone MPC independently delivering insulin and pramlintide.

glucose below 70 mg/dL based on their carbohydrate ratio.

## 3. RESULTS

We evaluated several approaches at delivering pramlintide during AID. For all simulations, we used the OHSU T1D simulator (Resalat, et al. 2019) with the addition of the population PK/PD model of pramlintide incorporated. No meal announcements were provided to any of the virtual patients such that we relied on the meal detection and dosing algorithm for all meal insulin. We delivered real-world meals to the virtual patients and delivery times from the open source type 1 diabetes in exercise initiative data set (Riddell, et al. 2023) which includes carbohydrate meal amounts confirmed through food photography. These real-world meals scenarios lasted between 3 and 12 days.

### *3.1 Fixed ratio vs. independent control of pramlintide*

First, we evaluated the performance of the insulin-only MPC. Next we evaluated the MPC when pramlintide was dosed at a fixed ratio of 6 mcg of pramlintide per unit of insulin. This delivery approach was designed to emulate an AID that delivers a coformulation of insulin and pramlintide at a ratio of 6:1. We evaluated this fixed ratio delivery of insulin for an MPC that had a process model that was unaware of the pramlintide delivery (pram-unaware), and also with an MPC that was aware of the pramlintide delivery (pram-aware). Finally, we evaluated a dual-hormone MPC that allowed for independent delivery of insulin and pramlintide.

Results from these experiments are shown in Table 3. The percent time in range (TIR, 70-180 mg/dL) improved from 64.1% in the insulin-only experiment to 91.6% when insulin and pramlintide were delivered at a fixed ratio with the process model in the MPC aware of the pramlintide being delivered. There was significant improvement in TIR when the process model was aware of the pramlintide (78.7% vs 91.6%). Interestingly, the TIR was higher for the insulin and pramlintide delivered as a fixed ratio compared with independent control of the insulin and pramlintide (91.6% vs. 87.5%). While this result was surprising, we surmised that was because there was a benefit of having pramlintide on board at the time that the meal was consumed. Since the MPC had no knowledge in advance of a meal being consumed, it was not possible to dose pramlintide in advance of the meal when pramlintide and insulin were dosed independently. Therefore, the pramlintide on board at the time of the meal detection was oftentimes close to zero for the independent control of pramlintide and insulin. An example from one participant across each of these conditions is shown in Figure 2 in response to a meal. Notice that the peak CGM is significantly lower when pramlintide is delivered compared with the insulin-only example. Notice also that when the MPC process model is unaware of the pramlintide, it is not as successful at bringing the person's glucose down to baseline following the meal because the process model is not aware that the carbohydrate will be on board for a longer period of time due to the delayed gastric emptying caused by the pramlintide.

| | **Insulin only** | **Insulin + pram 6:1 fixed ratio Pram unaware** | **Ins + pram 6:1 ratio Pram aware** | **Ins + pram Independent control** |
|---|---|---|---|---|
| Very low (<54 mg/dL) | 0.001 (0.11) | 0 (0) | 0 (0) | 0.01 (0.03) |
| Low (< 70 mg/dL) | 0.08 (0.13) | 0.01 (0.04) | 0.11 (0.22) | 0.09 (0.20) |
| In range (70-180 mg/dL) | 64.13 (12.46) | 78.71 (12.41) | 91.59 (8.58) | 87.50 (9.30) |
| High (>180 mg/dL) | 35.79 (12.5) | 21.28 (12.40) | 8.3 (8.5) | 12.40 (9.20) |
| Very high (>250 mg/dL) | 10.38 (8.7) | 2.26 (3.8) | 0.50 (1.7) | 1.20 (2.30) |
| LBGI | 0.10 (0.06) | 0.06 (0.04) | 0.10 (0.08) | 0.10 (0.08) |
| HBGI | 8.36 (3.78) | 4.76 (2.24) | 2.76 (1.41) | 3.37 (1.61) |
| Mean CGM | 171.44 (19.3) | 153.94 (13.10) | 141.17 (9.16) | 144.36 (9.47) |
| Low Events / Day | 0.06 (0.09) | 0.02 (0.05) | 0.1 (0.2) | 0.08 (0.16) |
| Total daily insulin (U) | 39.22 (15.9) | 42.04 (17.05) | 43.80 (17.94) | 43.35 (17.69) |
| Total daily pram (mcg) | 0 (0) | 251.86 (102.3) | 262.33 (107.61) | 269.01 (116.79) |
| MDD Sensitivity | 0.51 (0.14) | 0.18 (0.12) | 0.20 (0.10) | 0.21 (0.13) |
| MDD FP/day | 0.81 (0.39) | 1.31 (0.54) | 1.5 (0.60) | 1.53 (0.56) |
| MDD Units/day | 7.08 (3.6) | 4.92 (3.23) | 4.75 (2.74) | 4.90 (2.83) |

Table 3: Results comparing an insulin-only MPC with insulin and pramlintide delivered as a fixed ratio of 6 mcg pramlintide : 1 u insulin and when delivered independently.

The interquartile plot in Figure 3 shows how across 99 virtual patients, the pramlintide-aware MPC delivering at a 6:1 fixed ratio significantly reduced the postprandial glucose spike.

The meal detection and dosing (MDD) algorithm detected 51% of the meals in the insulin-only study with about 0.81 false positives per day. We defined a true positive meal detection if the meal was detected within 45 minutes following a meal consumption. The sensitivity of the MDD algorithm dropped because the slope of the CGM following a meal was lower and also because the meal detection was likely detected beyond the 45-minute dosing window.

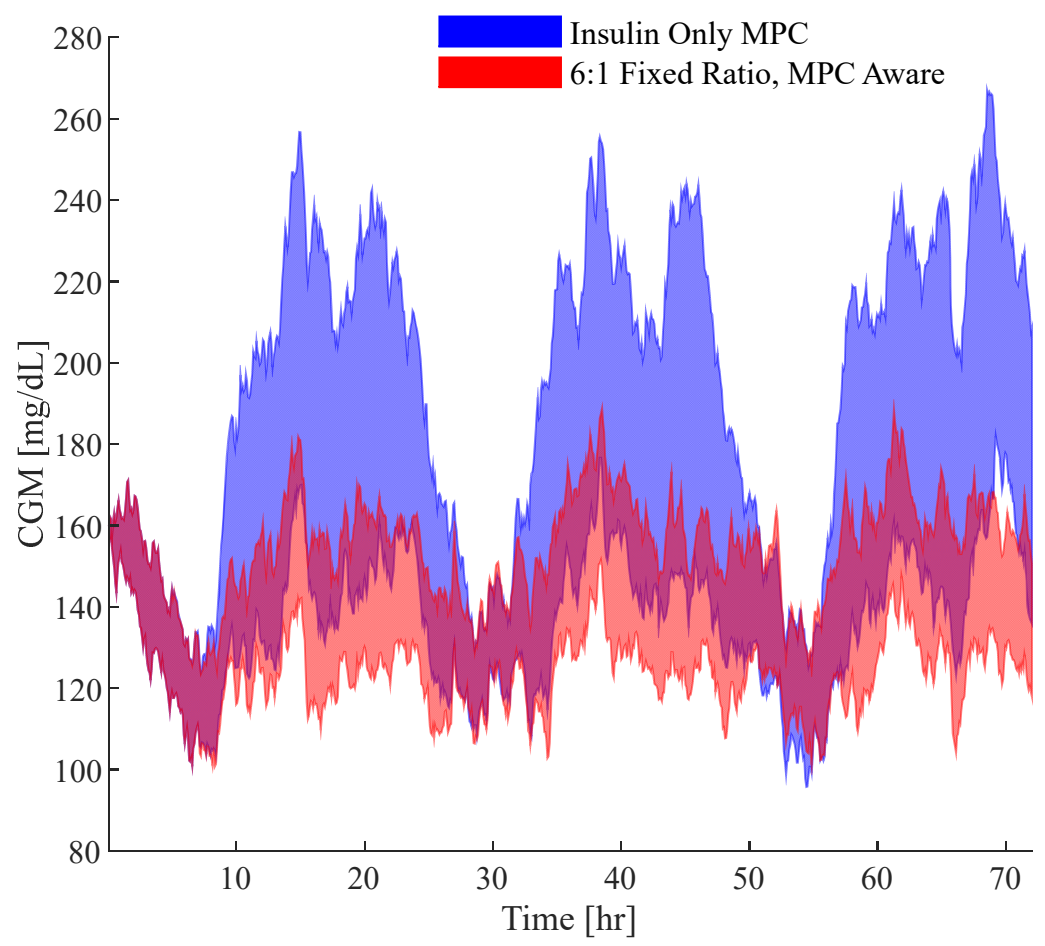

Figure 3: Interquartile plot results across 6-day simulations from all participants for the insulin-only MPC (blue) and the AID with pramlintide delivered as a 6:1 fixed ratio (red).

### *3.2 Pramlintide delivery at constant basal levels*

Pramlintide has the potential to cause nausea. It may be that delivering high doses of pramlintide is what causes the nausea and that if pramlintide were delivered at a constant rate throughout the day, there may be a benefit without the nausea. We evaluated how different constant basal pramlintide delivery amounts impact glucose outcomes in a pramlintide-aware MPC. We delivered pramlintide at a constant rate that was a function of the participant's average daily insulin infusion rate (IIR) such that basal pramlintide was delivered at either 2, 4, 6, 8, or 10 mcg/hr × IIR. Results shown in Table 4 indicate that TIR improved as more basal pramlintide was delivered compared with the insulin-only arm. The best TIR (89.1%) was achieved for the 10 mcg/hr × IIR delivery rate. This was worse performance than when insulin and pramlintide were delivered as a fixed-ratio with a pramlintide-aware MPC, even though the average total daily pramlintide delivered was more during the constant basal pramlintide delivery.

Supplementary Figure 1 in the arxiv Supplementary Materials (Jacobs et al. 2025) shows results from a single subject that shows how the postprandial peak glucose drops as more basal pramlintide is delivered at a constant level.

### *3.3 Evaluation in humans*

We evaluated the pramlintide unaware MPC in a human study (NCT06422325, IRB25279) whereby participants came to the OHSU clinic and consumed two meals either using an insulin-only MPC or a pramlintide unaware MPC that delivered pramlintide at a fixed ratio of 6 mcg pramlintide : 1 u of insulin. Participants consumed approximately 60 g of carbohydrates for each meal that was a function of their body weight and total daily carbohydrate needs. We performed a preliminary safety analysis on the first several subjects. Results from one of these participants is shown in Figure 4. Postprandial glucose is significantly reduced with pramlintide.

| | Insulin only | Basal pram 2 mcg/hr × Basal IIR | Basal pram 4 mcg/hr × Basal IIR | Basal pram 6 mcg/hr × Basal IIR | Basal pram 8 mcg/hr × Basal IIR | Basal pram 10 mcg/hr × Basal IIR |
|---|---|---|---|---|---|---|
| Very low (<54 mg/dL) | 0 (0.01) | 0 (0.01) | 0 (0.02) | 0.01 (0.08) | 0.01 (0.06) | 0.02 (0.09) |
| Low (< 70 mg/dL) | 0.08 (0.13) | 0.12 (0.20) | 0.09 (0.14) | 0.10 (0.21) | 0.11 (0.26) | 0.13 (0.28) |
| In range (70-180 mg/dL) | 64.13 (12.46) | 70.86 (12.57) | 78.39 (11.88) | 83.47 (11.17) | 86.90 (10.48) | 89.05 (9.79) |
| High (>180 mg/dL) | 35.79 (12.47) | 29.01 (12.57) | 21.52 (11.86) | 16.43 (11.14) | 12.99 (10.43) | 10.83 (9.73) |
| Very high (>250 mg/dL) | 10.38 (8.67) | 5.82 (6.90) | 3.21 (5.60) | 2.05 (4.39) | 1.32 (3.17) | 0.94 (2.67) |
| LBGI | 0.10 (0.06) | 0.11 (0.06) | 0.10 (0.06) | 0.11 (0.07) | 0.12 (0.09) | 0.12 (0.09) |
| HBGI | 8.36 (3.78) | 6.49 (3.17) | 5.00 (2.67) | 4.06 (2.25) | 3.46 (1.93) | 3.08 (1.73) |
| Mean CGM | 171.44 (19.28) | 162.16 (16.71) | 153.99 (14.33) | 148.28 (12.64) | 144.54 (11.23) | 142.13 (10.43) |
| Low Events / Day | 0.06 (0.09) | 0.10 (0.13) | 0.09 (0.11) | 0.08 (0.12) | 0.09 (0.13) | 0.09 (0.15) |
| Total daily insulin (U) | 39.22 (15.94) | 40.64 (16.98) | 41.89 (17.69) | 42.71 (17.90) | 43.26 (18.00) | 43.59 (18.00) |
| Total daily pram (mcg) | 0 (0) | 56.35 (21.97) | 112.70 (43.94) | 169.05 (65.91) | 225.39 (87.88) | 281.74 (109.85) |
| MMD Sensitivity | 0.51 (0.14) | 0.33 (0.16) | 0.25 (0.15) | 0.20 (0.14) | 0.17 (0.12) | 0.15 (0.11) |
| MMD FP/day | 0.81 (0.39) | 1.39 (0.51) | 1.53 (0.50) | 1.56 (0.50) | 1.54 (0.47) | 1.54 (0.50) |
| MMD Units/day | 7.08 (3.63) | 6.83 (3.74) | 5.94 (2.93) | 5.16 (2.64) | 4.61 (2.42) | 4.29 (2.32) |

Table 4: Mean and standard deviation results comparing an insulin-only MPC with and insulin-only MPC that is aware of the pramlintide delivery and the PK/PD of pramlintide, but when pramlintide is delivered as a fixed basal amount constantly throughout the day and night at different levels as a function of the basal insulin infusion rate (IIR): 2, 4, 6, 8, and 10 mcg/hr times the IIR.

## 4. DISCUSSION

We present a new model that describes PK/PD of pramlintide within a glucoregulatory model. We demonstrate that including this model within the process model of an MPC AID algorithm significantly improves glucose outcomes. While it is important to inform the MPC model of the pramlintide delivery using the PK/PD model of pramlintide, there is not a benefit of independent control of insulin and pramlintide (Table 3). Delivering constant basal pramlintide as a function of daily insulin requirements improves time in range compared with an insulin only MPC. The benefit was not as significant as when pramlintide was delivered as a fixed ratio with insulin that also dosed automated meal insulin using MDD.

It was surprising that results did not show a very high time below range (70 mg/dL) as is oftentimes observed in people with T1D on AID therapy. This was likely because there were no meal announcements given in this study, meaning that participants' glucose levels were generally high (Supplemental Figure 1 in Jacobs et al. 2025).

Prior work has been done on pramlintide PK/PD including a model proposed by Ramkissoon et al. (2014) and a simpler model by Furió-Novejarque and Bondia (2024). The architecture of the model proposed in the current manuscript is simpler than that proposed by Ramkissoon et al. For example, additional unobserved PK compartments that move pramlintide out of the plasma and then back into the plasma in the Ramkissoon model were not found to be necessary. Furió-Novejarque et al. also identified a simpler model than the

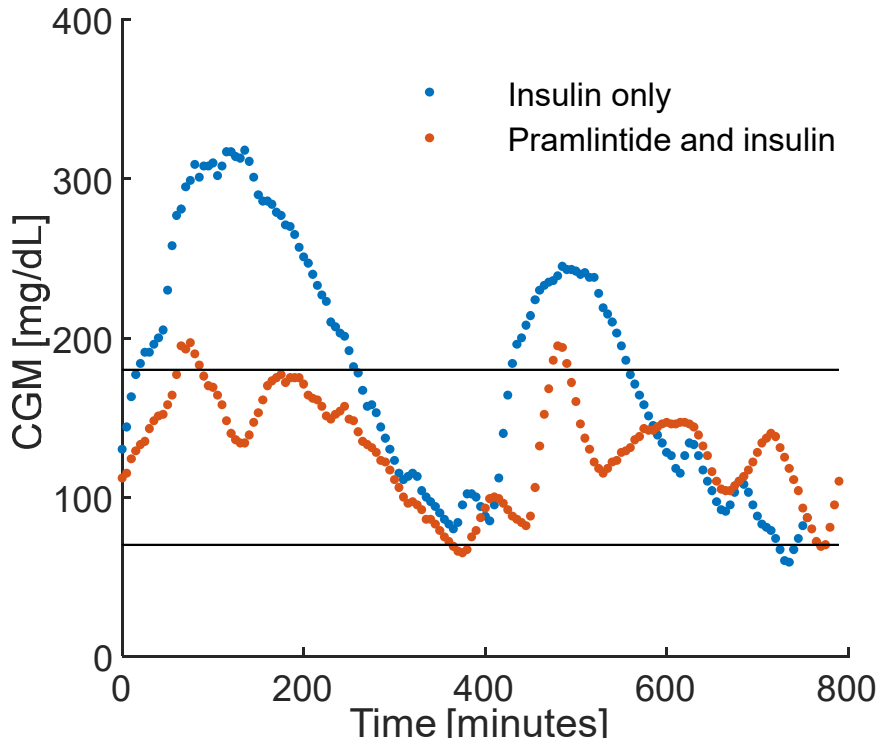

Figure 4: Results from one subject in ongoing study evaluating the OHSU insulin+pramlintide MPC vs. insulin-only. Postprandial time in range is nearly 100% when pramlintide is delivered.

Ramkissoon model. The best Furió-Novejarque et al. model was different than the one we present both in terms of the PK and also the PD, likely due to our use of intact and active pramlintide metabolite measurements for model fitting.

Limitations of this paper are that the primary findings are from *in silico* simulations, though human results are also shown.

## 5. CONCLUSIONS

Pramlintide can improve glucose outcomes in AID therapy. A coformulation of insulin and pramlintide at a ratio of 6:1 is a promising way to enable automated insulin and pramlintide delivery using current insulin pump technology. Independent delivery of insulin and pramlintide and constant basal delivery of pramlintide also provides benefit. Future work will include training the meal detection neural network on CGM data collected during insulin and pramlintide delivery to improve accuracy.

## ACKNOWLEDGEMENTS

Work supported by NIH/NIDDK 1R01DK129382-01. PGJ and LMW receive research support from Dexcom and Eli Lilly. PGJ is a shareholder and co-founder of Pacific Diabetes Technologies which may have interest in results.

## SUPPLEMENTARY MATERIALS

Supplemental Figure 1 below shows CGM traces for a single in silico subject in response to a meal when the person is receiving different rates of constant basal pramlintide.

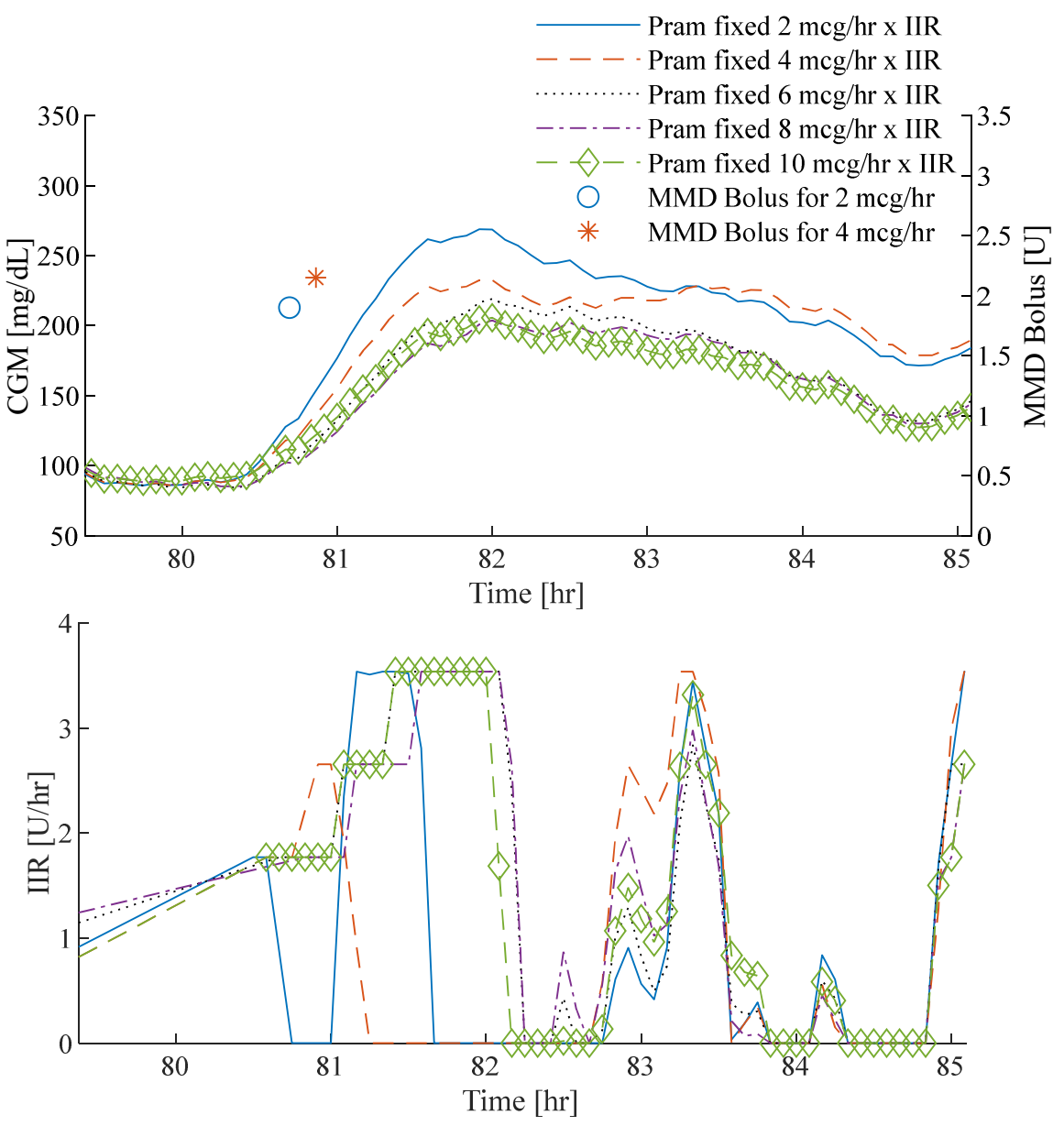

**Supplemental Figure 1:** Results from *in silico* subject consuming the same meal using an insulin-only MPC with pramlintide delivered at 5 different rates as a multiple of the person's automated basal insulin infusion rate (IIR): 2 mcg/hr × IIR, 4 mcg/hr × IIR, 6 mcg/hr × IIR, 8 mcg/hr × IIR, 10 mcg/hr × IIR. Meal detection insulin is dosed later as more basal pramlintide is dosed as shown by the blue circle and red

### Candidate models

We evaluated a number of models, differeing somewhat in structure, towards the goal of designating a preferred model which we will apply to the problem of glucose control. The model evaluation problem was divided into three parts, corresponding to the insulin pharmacokinetics (PK), pramlintide PK, and glucoregulatory pharmacodynamics (PD) subsystems. We evaluted six insulin PK, six pramlintide PK, and eight glucoregulatory PD models.

For each subsystem, best-fitting parameters were found by random sampling (implemented in Stan, as described further below). For each set of parameter values, the system of ODEs for the subsystem was solved, and the estimated values of observable quantities (insulin for insulin PK, intact pramlintide and active pramlintide metabolite for pramlintide PK, and glucose for the glucoregulatory PD) compared with experimental measurements.

### Candidate models for insulin PK

We studied six models for insulin pharmacokinetics as shown in Fig. 1. Compartments $Q_{i1}$ , $Q_{i1a}$ , and $Q_{i1b}$ represent a subcutaneous bolus. LD, LDa, and LDb represent local degradation with Michaelis-Menten kinetics. Compartment $Q_{i2}$ represents transport from the subcutaneous region to plasma, and Qi3 represents insulin in plasma. Compartment e is an elimination compartment. 3 The variations which were evaluated are whether there is one pathway or two pathways, in parallel, from subcutaneous bolus to plasma, whether there is local degradation in the one-pathway model, and whether the rate coefficient from the intermediate compartment $Q_{i2}$ to plasma $Q_{i3}$ is an independent free parameter. As with the pramlintide PK models, for each model structure there is a corresponding set of differential equations. For each model in which local degradataion (LD) is present, the parameters $K_M$ and $V_{max}$ have the values $K_M$ = 62.6 mU and $V_{max}$ = 1.93 mU/min, respectively, and m is body mass in kg.

- Insulin PK models 1 and 2. Model 1 is the same as model 2, except that model 1 has $k_{i2}=k_{i1}$.

$$
\begin{aligned}
\dot{Q}_{i_1} &= -k_{i_1} Q_{i_1} \\
\dot{Q}_{i_2} &= k_{i_1} Q_{i_1} - k_{i_2} Q_{i_2} \\
\dot{Q}_{i_3} &= k_{i_2} Q_{i_2} - k_e Q_{i_3}
\end{aligned}
$$

- Insulin PK models 3 and 4. Model 3 is the same as model 4, except model 3 has $k_{i2}=k_{i1}$.

$$
\mathrm{LD} = \frac{V_{\max}/m}{K_M/m + Q_{i_1}} Q_{i_1}
$$

$$
\begin{aligned}
\dot{Q}_{i_1} &= -k_{i_1} Q_{i_1} - \mathrm{LD} \\
\dot{Q}_{i_2} &= k_{i_1} Q_{i_1} - k_{i_2} Q_{i_2} \\
\dot{Q}_{i_3} &= k_{i_2} Q_{i_2} - k_e Q_{i_3}
\end{aligned}
$$

- Insulink PK models 5 and 6. Model 5 I shte same as model 6, except model 5 has $k_{i3}=k_{i1}$.

$$
\mathrm{LD}_a = \frac{V_{\max}/m}{K_M/m + Q_{i_{1a}}} Q_{i_{1a}}
$$

$$
\mathrm{LD}_b = \frac{V_{\max}/m}{K_M/m + Q_{i_{1b}}} Q_{i_{1b}}
$$

$$
\begin{aligned}
\dot{Q}_{i_{1a}} &= -k_{i_1} Q_{i_{1a}} - \mathrm{LD}_a \\
\dot{Q}_{i_{1b}} &= -k_{i_2} Q_{i_{1b}} - \mathrm{LD}_b \\
\dot{Q}_{i_2} &= k_{i_1} Q_{i_{1a}} - k_{i_3} Q_{i_2} \\
\dot{Q}_{i_3} &= k_{i_3} Q_{i_2} + k_{i_2} Q_{i_{1b}} - k_e Q_{i_3}
\end{aligned}
$$

Candidate model diagrams for the insulin PK are shown below in Figure 1.

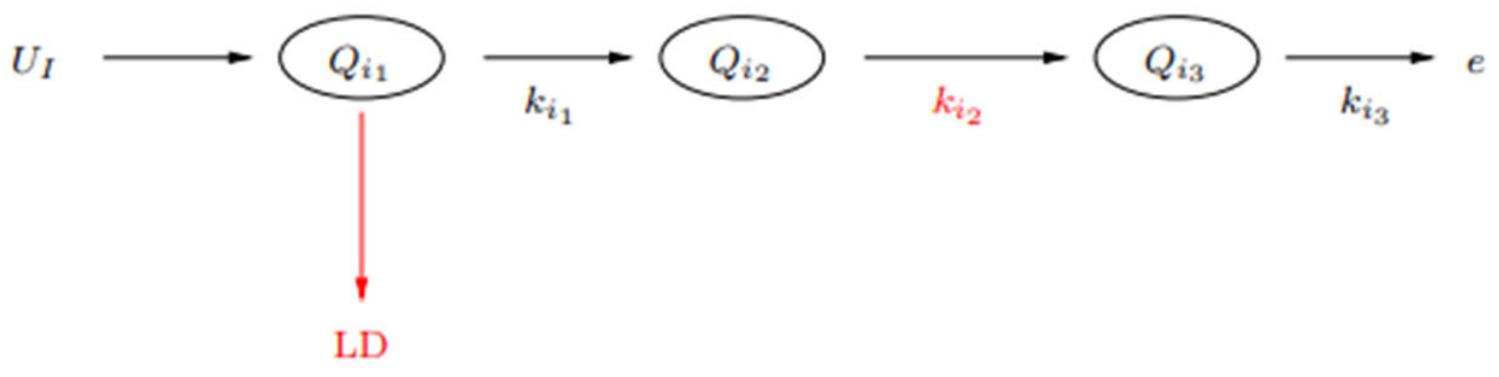

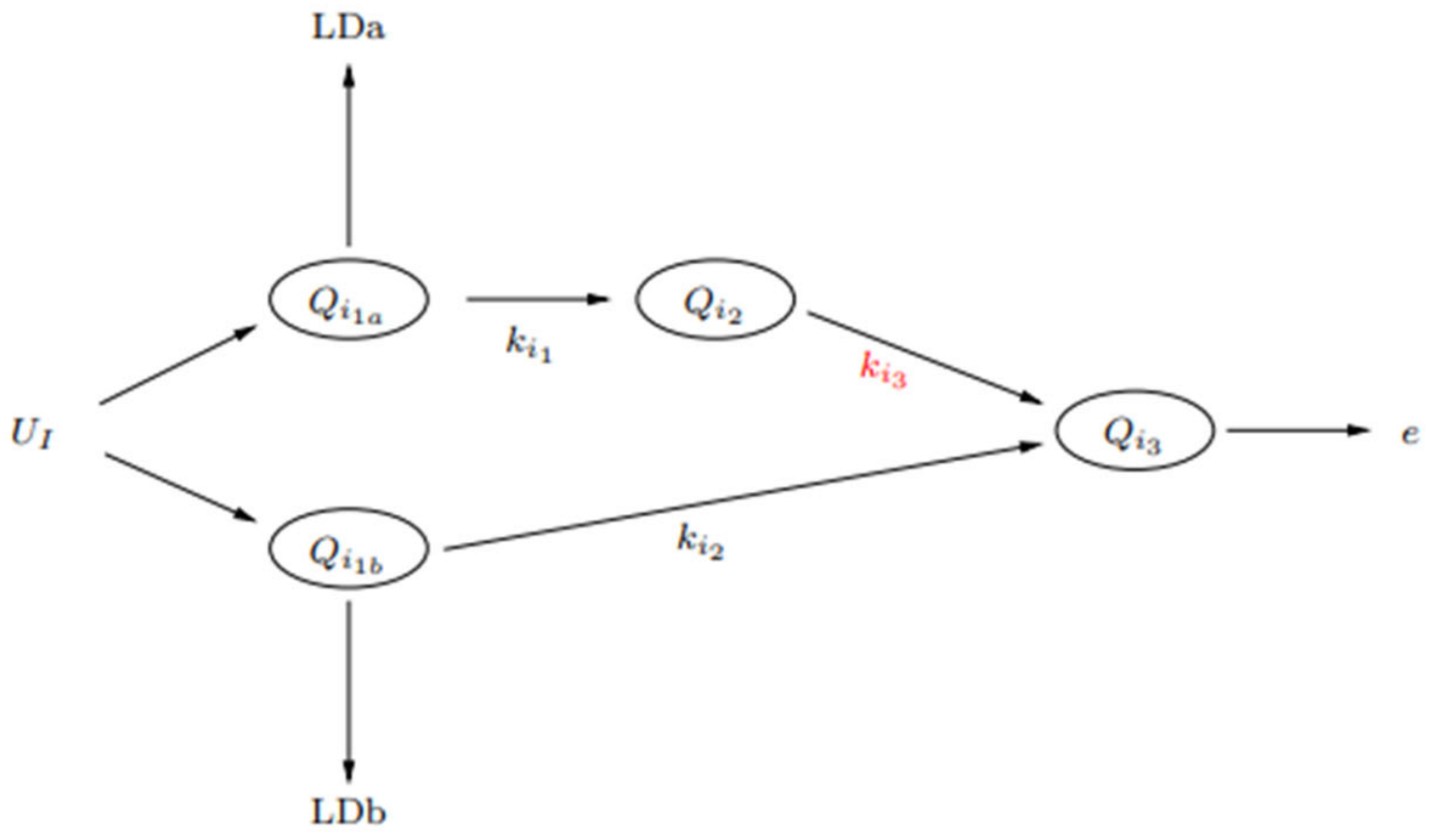

Supplemental Figure 2: Candicdate models for insulin PK. Above models 1, 2, 3, and 4 with one pathway and with or without local degradation. Below, models 5 and 6, with two pathways and local degradation. Model 1: ki2 = ki1 , no local degradation. Model 2: $k_{i2} \neq k_{i1}$ , no local degradation. Model 3: $k_{i2} = k_{i1}$, with local degradation. Model 4: $k_{i2} \neq k_{i1}$ , with local degradation. Model 5: $k_{i3} = k_{i1}$. Model 6: $k_{i3} \neq k_{i1}$.

**Candidate models for pramlintide PK**

We studied six models for pramlintide pharmacokinetics as shown in Fig. 2. Compartment $Q_{P1}$ represents a pramlintide bolus delivered subcutaneously, $Q_{P2}$ represents intact pramlintide in plasma, and $Q_{P3}$ represents pramlintide active metabolite (des-lys pramlintide) in plasma. $Q_{e2}$ and $Q_{e3}$ are elimination compartments. The variations which were evaluated are the presence or absence of a path from compartment $Q_{P1}$ to $Q_{P3}$ , and from compartment $Q_{P2}$ to $Q_{P3}$ , and the presence or absence of so-called hidden compartments $Q_{P4}$ and $Q_{P5}$ , which are not directly connected to an input or elimination compartment. For each model structure, there are similar sets of differential equations.

- Pramlintide PK model 1

$$\begin{aligned}\dot{Q}_{P_1} &= -k_{P_2,P_1}Q_{P_1} - k_{P_3,P_1}Q_{P_1}\\ \dot{Q}_{P_2} &= k_{P_2,P_1}Q_{P_1} - k_{e_2,P_2}Q_{P_2}\\ \dot{Q}_{P_3} &= k_{P_3,P_1}Q_{P_1} - k_{e_3,P_3}Q_{P_3}\end{aligned}$$

- Pramlintide PK model 2

$$\begin{aligned}\dot{Q}_{P_1} &= -k_{P_2,P_1}Q_{P_1} - k_{P_3,P_1}Q_{P_1}\\ \dot{Q}_{P_2} &= k_{P_2,P_1}Q_{P_1} - k_{P_3,P_2}Q_{P_2} - k_{e_2,P_2}Q_{P_2}\\ \dot{Q}_{P_3} &= k_{P_3,P_1}Q_{P_1} + k_{P_3,P_2}Q_{P_2} - k_{e_3,P_3}Q_{P_3}\end{aligned}$$

- Pramlintide PK model 3

$$
\begin{aligned}
\dot{Q}_{P_1} &= -k_{P_2,P_1}Q_{P_1} \\
\dot{Q}_{P_2} &= k_{P_2,P_1}Q_{P_1} - k_{P_3,P_2}Q_{P_2} - k_{e_2,P_2}Q_{P_2} \\
\dot{Q}_{P_3} &= k_{P_3,P_2}Q_{P_2} - k_{e_3,P_3}Q_{P_3}
\end{aligned}
$$

- Pramlintide PK model 4

$$
\begin{aligned}
\dot{Q}_{P_1} &= -k_{P_2,P_1}Q_{P_1} - k_{P_3,P_1}Q_{P_1} \\
\dot{Q}_{P_2} &= k_{P_2,P_1}Q_{P_1} - k_{e_2,P_2}Q_{P_2} - k_{P_4,P_2}Q_{P_2} + k_{P_2,P_4}Q_{P_4} \\
\dot{Q}_{P_3} &= k_{P_3,P_1}Q_{P_1} - k_{e_3,P_3}Q_{P_3} - k_{P_5,P_3}Q_{P_3} + k_{P_3,P_5}Q_{P_5} \\
\dot{Q}_{P_4} &= k_{P_4,P_2}Q_{P_2} - k_{P_2,P_4}Q_{P_4} \\
\dot{Q}_{P_5} &= k_{P_5,P_3}Q_{P_3} - k_{P_3,P_5}Q_{P_5}
\end{aligned}
$$

- Pramlintide PK model 5

$$
\begin{aligned}
\dot{Q}_{P_1} &= -k_{P_2,P_1}Q_{P_1} - k_{P_3,P_1}Q_{P_1} \\
\dot{Q}_{P_2} &= k_{P_2,P_1}Q_{P_1} - k_{P_3,P_2}Q_{P_2} - k_{e_2,P_2}Q_{P_2} - k_{P_4,P_2}Q_{P_2} + k_{P_2,P_4}Q_{P_4} \\
\dot{Q}_{P_3} &= k_{P_3,P_1}Q_{P_1} + k_{P_3,P_2}Q_{P_2} - k_{e_3,P_3}Q_{P_3} - k_{P_5,P_3}Q_{P_3} + k_{P_3,P_5}Q_{P_5} \\
\dot{Q}_{P_4} &= k_{P_4,P_2}Q_{P_2} - k_{P_2,P_4}Q_{P_4} \\
\dot{Q}_{P_5} &= k_{P_5,P_3}Q_{P_3} - k_{P_3,P_5}Q_{P_5}
\end{aligned}
$$

- Pramlintide PK model 6

$$
\begin{aligned}
\dot{Q}_{P_1} &= -k_{P_2,P_1}Q_{P_1} \\
\dot{Q}_{P_2} &= k_{P_2,P_1}Q_{P_1} - k_{P_3,P_2}Q_{P_2} - k_{e_2,P_2}Q_{P_2} - k_{P_4,P_2}Q_{P_2} + k_{P_2,P_4}Q_{P_4} \\
\dot{Q}_{P_3} &= k_{P_3,P_2}Q_{P_2} - k_{e_3,P_3}Q_{P_3} - k_{P_5,P_3}Q_{P_3} + k_{P_3,P_5}Q_{P_5} \\
\dot{Q}_{P_4} &= k_{P_4,P_2}Q_{P_2} - k_{P_2,P_4}Q_{P_4} \\
\dot{Q}_{P_5} &= k_{P_5,P_3}Q_{P_3} - k_{P_3,P_5}Q_{P_5}
\end{aligned}
$$

Candidate pramlintide PK model diagrams are shown in Figure 2 below.

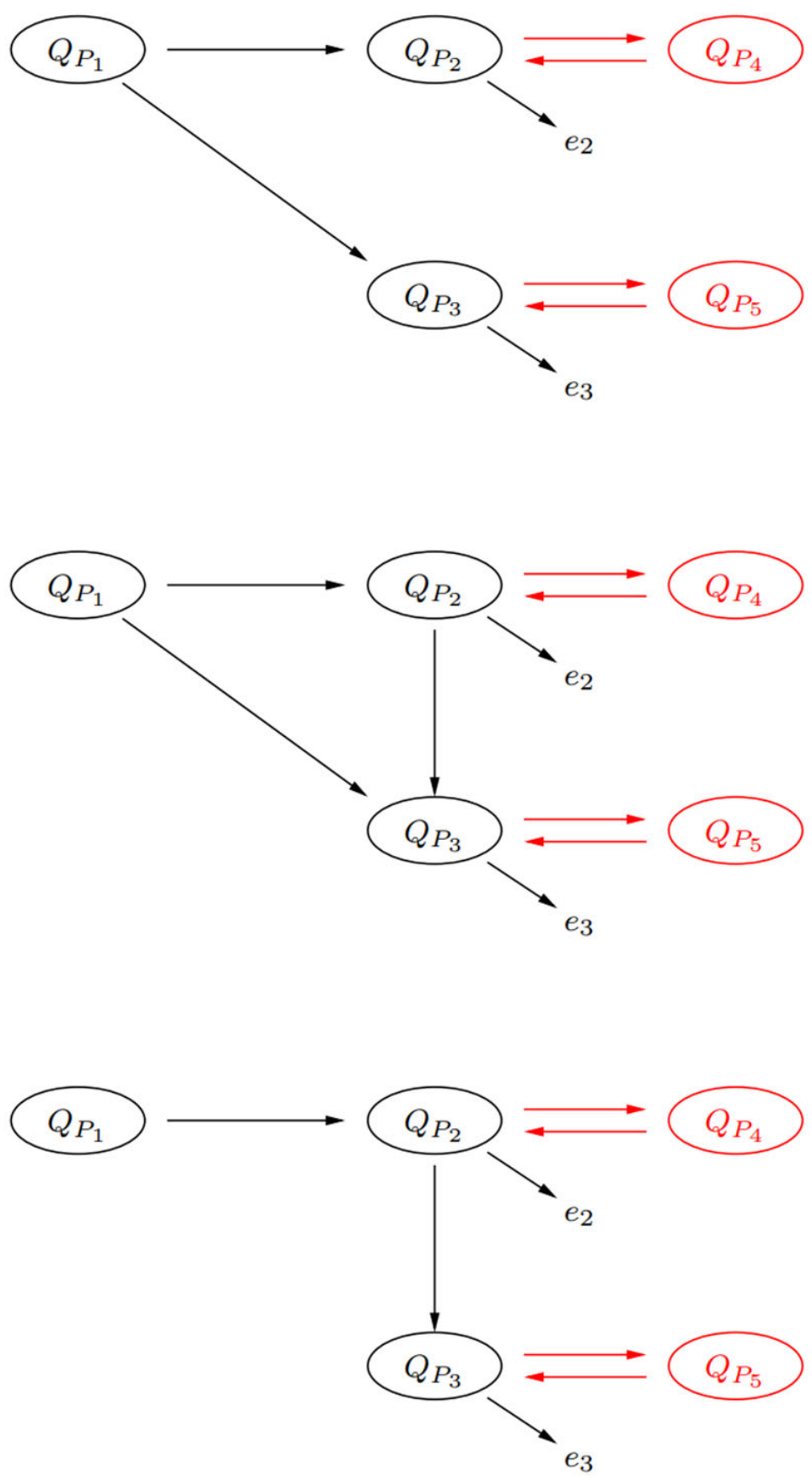

Supplemental Figure 3: Candidate models for pramlintide pharmacokinetics. Top, models 1 and 4, with pathway $Q_{P1} \rightarrow Q_{P3}$ , and without pathway $Q_{P2} \rightarrow Q_{P3}$ . Center, models 2 and 5, with pathways $Q_{P1} \rightarrow Q_{P3}$ and $Q_{P2} \rightarrow Q_{P3}$ . Bottom, models 3 and 6, without pathway $Q_{P1} \rightarrow Q_{P3}$ , and with pathway $Q_{P2} \rightarrow Q_{P3}$ . Model 1: with pathway $Q_{P1} \rightarrow Q_{P3}$ , and without pathway $Q_{P2} \rightarrow Q_{P3}$ , without hidden compartments. Model 2: with pathways $Q_{P1} \rightarrow Q_{P3}$ and $Q_{P2} \rightarrow Q_{P3}$ , without hidden compartments. Model 3: without pathway $Q_{P1} \rightarrow Q_{P3}$ , and with pathway $Q_{P2} \rightarrow Q_{P3}$ , without hidden compartments. Model 4: with pathway $Q_{P1} \rightarrow Q_{P3}$ , and without pathway $Q_{P2} \rightarrow Q_{P3}$ , with hidden compartments. Model 5: with pathways $Q_{P1} \rightarrow Q_{P3}$ and $Q_{P2} \rightarrow Q_{P3}$ , with hidden compartments. Model 6: without pathway $Q_{P1} \rightarrow Q_{P3}$ , and with pathway $Q_{P2} \rightarrow Q_{P3}$ , with hidden compartments..

**Candidate models for glucose PD**

We studied eight models for glucoregulatory pharmacodynamics as shown in Fig. 3. Compartment $Q_{S0}$ , or $Q_{S1}$ if $Q_{S0}$ is not present, represents carbohydrate input to the stomach. Compartment $Q_{S2}$ represents carbohydrates in the gastrointestinal system. Compartment $Q_{G1}$ represents glucose in plasma, while $Q_{G2}$ represents glucose in the interstitium. There is an input path in addition to carbohydrates in the digestive system, namely endogenous glucose production (EGP). There are also several output paths. The main glucose pathway leads from the carbohydrate compartments through glucose in plasma to the interstitium and thence to insulin-mediated glucose uptake (IMGU). There is an output pathway for carbohydrate elimination (SE), representing unutilized carbohydrates in the digestive system, and for glucose outputs, renal elimination (RE), and non-insulin-mediated glucose uptake (NIMGU), comprising mostly glucose uptake in the brain.

The equations for the glucoregulatory models are the following. The factor $\alpha$ appearing in the equations for $\dot{Q}_{G1}$ is the conversion factor for converting glucose to the correct units, $\alpha$=1000/180 mmol/g.

- *Glucoregulatory models 1, 2, 3, 4, 5, and 6:* Models 1, 2, and 3 are the same as model 4 with the following differences. Model 1: $k_{G1,S2} = k_{S2,S1}$ , $k_{SE}$ fixed. Model 2: $k_{G1,S2} \neq k_{S2,S1}$ , $k_{SE}$ fixed. Model 3: $k_{G1,S2} = k_{S2,S1}$ , $k_{SE}$ variable. Model 4: $k_{G1,S2} \neq k_{S2,S1}$ , $k_{SE}$ variable. Models 5 and 6 have the same structure as model 3, except that SfP1 $\neq$ SfP2 in both cases; model 5 has no effect of P on kSE, while model 6 does.

$$\dot{Q}_{S_1} = U_{Q_{S_1}} - K_{S_2,S_1}Q_{S_1}$$
$$\dot{Q}_{S_2} = K_{S_2,S_1}Q_{S_1} - K_{\mathrm{SE}}Q_{S_2} - K_{G_1,S_2}Q_{S_2}$$
$$\dot{Q}_{G_1} = \alpha K_{G_1,S_2}Q_{S_2} - X_1Q_{G_1} + K_{G_1,G_2}Q_{G_2} + EGP - F_R - F_{01}^c$$
$$\dot{Q}_{G_2} = X_1Q_{G_1} - K_{G_1,G_2}Q_{G_2} - X_2Q_{G_2}$$

- *Glucoregulatory models 7 and 8:* Model 7 is the same as model 8, except model 7 is without action of P on $k_{SE}$.

$$\dot{Q}_{S_0} = U_{Q_{S_0}} - K_{S_1,S_0}Q_{S_0}$$
$$\dot{Q}_{S_1} = K_{S_1,S_0}Q_{S_0} - K_{S_2,S_1}Q_{S_1}$$
$$\dot{Q}_{S_2} = K_{S_2,S_1}Q_{S_1} - K_{\mathrm{SE}}Q_{S_2} - K_{G_1,S_2}Q_{S_2}$$
$$\dot{Q}_{G_1} = \alpha K_{G_1,S_2}Q_{S_2} - X_1Q_{G_1} + K_{G_1,G_2}Q_{G_2} + EGP - F_R - F_{01}^c$$
$$\dot{Q}_{G_2} = X_1Q_{G_1} - K_{G_1,G_2}Q_{G_2} - X_2Q_{G_2}$$

The candidate glucoregulatory PD model diagrams are shown below.

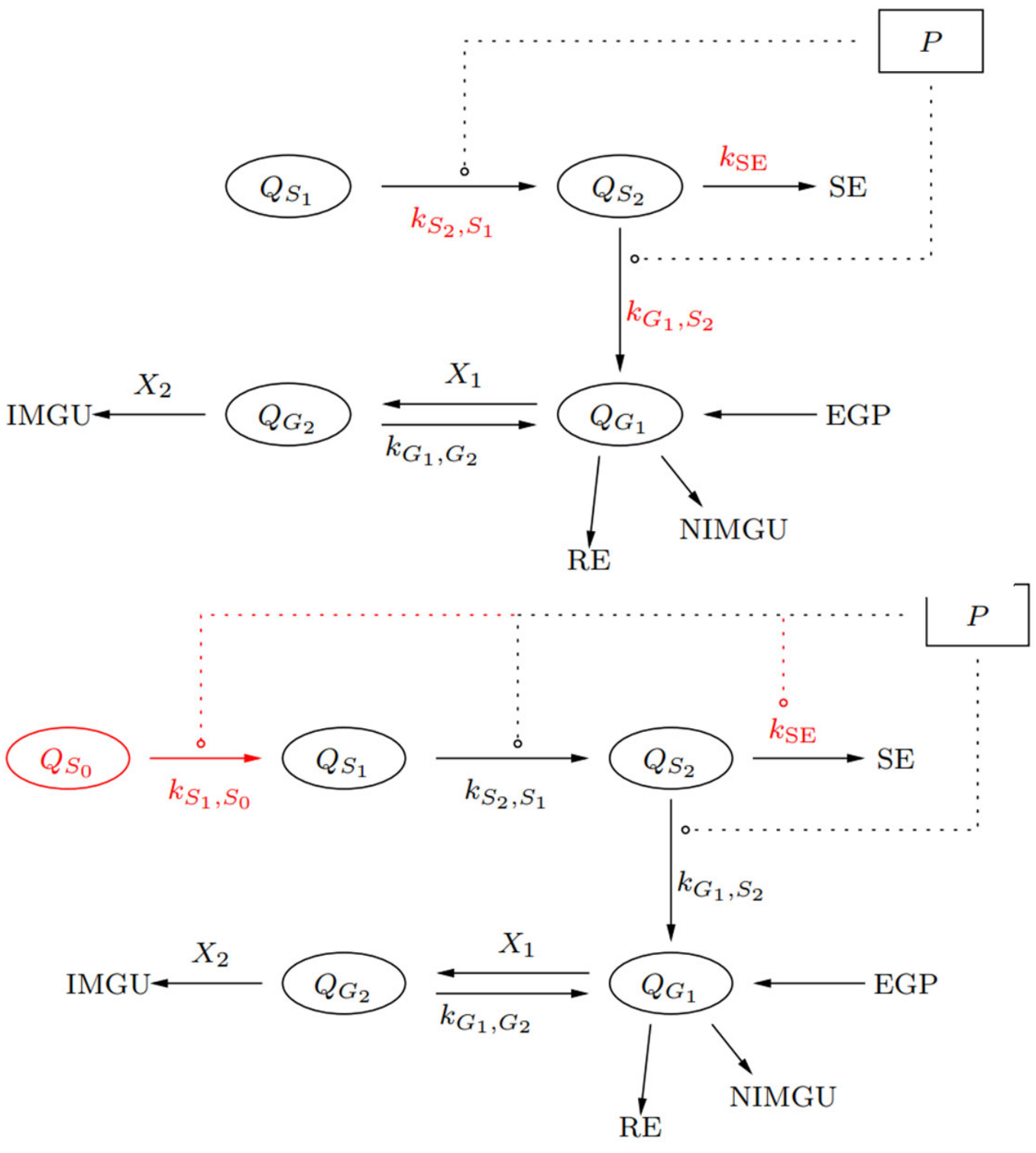

Supplemental Figure 4: Candidate models for glucoregulatory pharmacodynamics. Above, models 1, 2, 3, and 4, with $Sf_{P1} = Sf_{P2}$, without $Q_{S0}$, and without action of P on $k_{SE}$. Below, models 5, 6, 7, and 8, with $Sf_{P1} \neq Sf_{P2}$, with or without $Q_{S0}$, and with or without action of P on $k_{SE}$. Model 1: $k_{G1,S2} = k_{S2,S1}$, $k_{SE}$ fixed. Model 2: $k_{G1,S2} \neq k_{S2,S1}$, $k_{SE}$ fixed. Model 3: $k_{G1,S2} = k_{S2,S1}$, $k_{SE}$ variable. Model 4: $k_{G1,S2} \neq k_{S2,S1}$, $k_{SE}$ variable. Model 5: $Sf_{P1} \neq Sf_{P2}$, without $Q_{S0}$, and without action of P on $k_{SE}$. Model 6: $Sf_{P1} \neq Sf_{P2}$, without $Q_{S0}$, and with action of P on $k_{SE}$. Model 7: $Sf_{P1} \neq Sf_{P2}$, with $Q_{S0}$, and without action of P on $k_{SE}$. Model 8: $Sf_{P1} \neq Sf_{P2}$, with $Q_{S0}$, and with action of P on $k_{SE}$.

**Parameter inference via Markov Chain Monte Carlo**

For each candidate model, we searched for parameter values to maximize goodness of fit, defined as mean squared error (MSE), the square of the difference between the value predicted by the model at a point in time, and the measured value for the quantity predicted by the model; the predicted quantities are insulin concentration for the insulin PK model, intact pramlintide and pramlintide active metabolite concentrations for the pramlintide PK model, and glucose concentration for the glucoregulatory PD model.

*Maximum likelihood and formal Bayesian inference*

The software we applied for parameter inference, Stan, implements full Bayesian inference, taking both prior and likelihood into account. Stan samples from the posterior of the joint distribution of parameters given the observed data,

$$p(\alpha \mid x, y) = \frac{p(x, y \mid \alpha)p(\alpha)}{\int_{\Omega(\alpha)} p(x, y \mid \alpha)p(\alpha)d\alpha}$$

However, in all of the results reported here, the prior for each parameter is a distribution which is uniform over an interval, and the posterior is therefore proportional to just the likelihood and constrained to the support of the prior (the interval over which the prior is nonzero). In effect, then, all inferences are just constrained maximum likelihood.

*Markov Chain Monte Carlo (MCMC) and Hamiltonian Monte Carlo*

A central goal in any Bayesian analysis is the construction of the posterior distribution of model parameters. This requires explicit or implicit integration over the parameter space, and, in general, numerical methods are necessary, due to the difficulty of carrying out the required integrations symbolically. When the dimensionality of the parameter space is small (i.e., there are few parameters), direct integration via quadrature rules may be possible. However, the computational burden for quadrature rules grows exponentially with the number of dimensions. When there are any more than a very few (in practice, perhaps about three) dimensions, random sampling is much more efficient. In any number of dimensions, the error of a random sampling estimate decreases proportionally to $1/\sqrt{n}$, where n is the number of samples. Markov chain Monte Carlo is a class of related algorithms for random sampling, which have the feature in common that it is possible to sample from a function which is only proportional to the posterior density of the parameters, that is, it is not necessary to normalize the posterior in order to sample from it. Hamiltonian Monte Carlo is a particular variant of MCMC which makes use of an analogy to classical mechanics to 6 define abstract position and momentum variables. The HMC formulation allows for faster exploration of the parameter space

*Software for MCMC*

To carry out sampling for model parameters, we employed Stan [1], a widely-used open source implementation of Hamiltonian Monte Carlo and other sampling algorithms. We worked with Stan through the CmdStanR interface [3] for the R statistical software system [2], which provides functions to launch Stan sampling and collect and extract results. The following sampling parameters were applied for each modeling task.

- Pramlintide PK: 1000 warmup samples, 4000 samples post-warmup
- Insulin PK: 500 warmup samples, 4000 samples post-warmup
- Glucoregulatory PD: 500 warmup samples, 2000 samples post-warmup

$\hat{r}$ diagnostic for MCMC Sampling

To gain insight about the sampling process, we examined the so-called r̂ diagnostic, which is defined, for each sampled parameter, as a function of the ratio of within-chains variance to between-chains variance:

$$\hat{R} = \sqrt{1 - \frac{1}{N} + \frac{\sum_{m=1}^{M}(\bar{\theta}_m^{(\cdot)} - \bar{\theta}_{\cdot}^{(\cdot)})^2}{\sum_{m=1}^{M} s_m^2}}$$

Where $\bar{\theta}_m^{(\cdot)}$ and $\bar{\theta}^{(\cdot)}$ are the per-chain mean and overall mean, respectively, and $s_m^2$ is the per-chain sample variance.

For well-mixed chains, the mean of each chain will converge to the posterior mean of the parameter, and therefore the overall mean will likewise converge, and the numerator of the fraction in the equation for $\hat{r}$ will converge to zero. The sample variance for each chain will converge to the posterior variance of the parameter, so the fraction (between-chains variance)/(within-chains variance) converges to zero, and $\hat{r}$ converges to 1. On the other hand, when chains are sampling substantially different regions in the parameter space and not overlapping, the between-chains variance does not converge to zero, and $\hat{r}$ does not converge to 1. A value of $\hat{r} \leq 1.1$ is considered to indicate acceptable mixing; otherwise, the diagnostic criterion indicates a sampling problem.

*Model selection for PK and PD models*

The crit4erion for model comparisons was the root mean squared error of the estimated values. We compare models via paired sample t-tests. We used the Bonferroni correction to nominal alpha=0.05 significance by dividing by number of comparisons.

**Linearization of the model**

Below we show the process for linearizing the final model that is given in Equations 1-14 in the manuscript. These state equations are included below again.

$$
\begin{aligned}
\dot{Q}_1 &= \left(-X - p_1 - k_{2,1}\right) Q_1 + k_{1,2}\, Q_2 + \frac{K\, M_2}{\dfrac{Sf_{P_2}\left(Q_{P_2} + Q_{P_3}\right)}{V_P} + 1} \\
\dot{Q}_2 &= k_{2,1}\, Q_1 - k_{1,2}\, Q_2 \\
\dot{x}_I &= U_I - k_a\, x_I \\
\dot{I} &= \frac{k_a\, x_I}{V_I} - k_e\, I \\
\dot{X} &= p_3\, I - p_2\, X \\
\dot{M}_1 &= U_m - \frac{M_1}{t^G_{\max}\left(\dfrac{Sf_{P_1}\left(Q_{P_2} + Q_{P_3}\right)}{V_P} + 1\right)} \\
\dot{M}_2 &= \frac{M_1}{t^G_{\max}\left(\dfrac{Sf_{P_1}\left(Q_{P_2} + Q_{P_3}\right)}{V_P} + 1\right)} - \frac{M_2}{t^G_{\max}\left(\dfrac{Sf_{P_2}\left(Q_{P_2} + Q_{P_3}\right)}{V_P} + 1\right)} \\
\dot{Q}_{P_1} &= U_P - k_{P_2,P_1}\, Q_{P_1} \\
\dot{Q}_{P_2} &= k_{P_2,P_1}\, Q_{P_1} - k_{e_2,P_2}\, Q_{P_2} - k_{P_3,P_2}\, Q_{P_2} \\
\dot{Q}_{P_3} &= k_{P_3,P_2}\, Q_{P_2} - k_{e_3,P_3}\, Q_{P_3}
\end{aligned}
$$

whereby the 10 state variables are $Q_1$, $Q_2$, $x_I$ , $I$, $X$, $M_1$, $M_2$, $Q_{P1}$ , $Q_{P2}$ , and $Q_{P3}$ . The linearized right-hand side, $D_X F(X^*) X$, is

$$
\dot{Q}_1 = \left(-p_1 - k_{2,1} - X^*\right) Q_1 + k_{1,2}\, Q_2 - Q_1^*\, X + \frac{K\, M_2}{\dfrac{\left(Q_{P_3}^* + Q_{P_2}^*\right) Sf_{P_2}}{V_P} + 1} - \frac{K\, Sf_{P_2}\, M_2^*\, Q_{P_2}}{\left(\dfrac{\left(Q_{P_3}^* + Q_{P_2}^*\right) Sf_{P_2}}{V_P} + 1\right)^2 V_P} - \frac{K\, Sf_{P_2}\, M_2^*\, Q_{P_3}}{\left(\dfrac{\left(Q_{P_3}^* + Q_{P_2}^*\right) Sf_{P_2}}{V_P} + 1\right)^2 V_P}
$$

$$
\dot{Q}_2 = k_{2,1}\, Q_1 - k_{1,2}\, Q_2
$$

$$
\dot{X}_I = -\left(k_a\, x_I\right)
$$

$$
\dot{I} = \frac{k_a\, x_I}{V_I} - k_e\, I
$$

$$
\dot{X} = p_3\, I - p_2\, X
$$

$$
\dot{M}_1 = -\left(\frac{M_1}{\left(\dfrac{\left(Q_{P_3}^* + Q_{P_2}^*\right) Sf_{P_1}}{V_P} + 1\right) t^G_{\max}}\right) + \frac{Sf_{P_1}\, M_1^*\, Q_{P_2}}{\left(\dfrac{\left(Q_{P_3}^* + Q_{P_2}^*\right) Sf_{P_1}}{V_P} + 1\right)^2 V_P\, t^G_{\max}} + \frac{Sf_{P_1}\, M_1^*\, Q_{P_3}}{\left(\dfrac{\left(Q_{P_3}^* + Q_{P_2}^*\right) Sf_{P_1}}{V_P} + 1\right)^2 V_P\, t^G_{\max}}
$$

$$
\begin{aligned}
\dot{M}_2 = {} & \frac{M_1}{\left(\dfrac{\left(Q_{P_3}^* + Q_{P_2}^*\right) Sf_{P_1}}{V_P} + 1\right) t^G_{\max}} - \frac{M_2}{\left(\dfrac{\left(Q_{P_3}^* + Q_{P_2}^*\right) Sf_{P_2}}{V_P} + 1\right) t^G_{\max}} + \left(\frac{Sf_{P_2}\, M_2^*}{\left(\dfrac{\left(Q_{P_3}^* + Q_{P_2}^*\right) Sf_{P_2}}{V_P} + 1\right)^2 V_P\, t^G_{\max}} - \frac{Sf_{P_1}\, M_1^*}{\left(\dfrac{\left(Q_{P_3}^* + Q_{P_2}^*\right) Sf_{P_1}}{V_P} + 1\right)^2 V_P\, t^G_{\max}}\right) Q_{P_2} \\
& + \left(\frac{Sf_{P_2}\, M_2^*}{\left(\dfrac{\left(Q_{P_3}^* + Q_{P_2}^*\right) Sf_{P_2}}{V_P} + 1\right)^2 V_P\, t^G_{\max}} - \frac{Sf_{P_1}\, M_1^*}{\left(\dfrac{\left(Q_{P_3}^* + Q_{P_2}^*\right) Sf_{P_1}}{V_P} + 1\right)^2 V_P\, t^G_{\max}}\right)
\end{aligned}
$$

$$
\dot{Q}_{P_1} = -\left(k_{P_2,P_1}\, Q_{P_1}\right)
$$

$$
\dot{Q}_{P_2} = k_{P_2,P_1}\, Q_{P_1} + \left(-k_{e_2,P_2} - k_{P_3,P_2}\right) Q_{P_2}
$$

$$
\dot{Q}_{P_3} = k_{P_3,P_2}\, Q_{P_2} - k_{e_3,P_3}\, Q_{P_3}
$$

The constant terms, $F(X^*) - D_X F(X^*) X^*$, are

$$F(X^*) - D_X F(X^*) X^*$$

$$= \begin{bmatrix} \dfrac{K\, Q_{P_3}^*\, Sf_{P_2}\, M_2^*}{{E_{P^*}^{(2)}}^2\, V_P} + \dfrac{K\, Q_{P_2}^*\, Sf_{P_2}\, M_2^*}{{E_{P^*}^{(2)}}^2\, V_P} + Q_1^*\, X^* \\ 0 \\ U_I \\ 0 \\ 0 \\ -\left(\dfrac{Q_{P_3}^*\, Sf_{P_1}\, M_1^*}{{E_{P^*}^{(1)}}^2\, V_P\, t_{\max}^G}\right) - \dfrac{Q_{P_2}^*\, Sf_{P_1}\, M_1^*}{{E_{P^*}^{(1)}}^2\, V_P\, t_{\max}^G} + U_m \\ -\left(Q_{P_3}^* \left(\dfrac{Sf_{P_2}\, M_2^*}{{E_{P^*}^{(2)}}^2\, V_P\, t_{\max}^G} - \dfrac{Sf_{P_1}\, M_1^*}{{E_{P^*}^{(1)}}^2\, V_P\, t_{\max}^G}\right)\right) - Q_{P_2}^* \left(\dfrac{Sf_{P_2}\, M_2^*}{{E_{P^*}^{(2)}}^2\, V_P\, t_{\max}^G} - \dfrac{Sf_{P_1}\, M_1^*}{{E_{P^*}^{(1)}}^2\, V_P\, t_{\max}^G}\right) \\ U_P \\ 0 \\ 0 \end{bmatrix}$$

Whereby we have defined the terms $E_{P^*}^{(1)}$ and $E_{P^*}^{(2)}$ for brevity.

$$E_{P^*}^{(1)} = \frac{\left(Q_{P_3}^* + Q_{P_2}^*\right) Sf_{P_1}}{V_P} + 1$$

$$E_{P^*}^{(2)} = \frac{\left(Q_{P_3}^* + Q_{P_2}^*\right) Sf_{P_2}}{V_P} + 1$$